\documentclass[%
 reprint,
superscriptaddress,
 amsmath,amssymb,
 aps,
]{revtex4-2}

\usepackage{graphicx}
\usepackage{dcolumn}
\usepackage{amssymb}
\usepackage{amsmath,bm}
\usepackage{graphicx,color}
\usepackage{bbm}
\usepackage{newtxmath}
\usepackage{multirow}
\usepackage{xcolor}
\usepackage{hyperref}
\usepackage{ulem}
\usepackage{tabularx}
\usepackage{bm}
\newcommand{\ah}[1]{\textcolor{blue}{#1}}

\begin{document}

\preprint{APS/123-QED}

\title{Finite Elasticity of the Vertex Model and its Role in Rigidity of Curved Cellular Tissues. }

\author{Arthur Hernandez}
\email{arthurhernandez@ucsb.edu}
\affiliation{Department of Physics, University of California Santa Barbara, Santa Barbara, CA 93106}

\author{Michael F. Staddon}
\affiliation{Center for Systems Biology Dresden, Dresden, Germany}
\affiliation{Max Planck Institute for the Physics of Complex Systems, Dresden, Germany}
\affiliation{Max Planck Institute of Molecular Cell Biology and Genetics, Dresden, Germany}

\author{Michael Moshe}
\affiliation{Racah Institute of Physics, The Hebrew University of Jerusalem, Jerusalem, Israel 91904}

\author{M. Cristina Marchetti}
\affiliation{Department of Physics, University of California Santa Barbara, Santa Barbara, CA 93106}

\date{\today}

\begin{abstract}
Using a mean field approach and simulation, we study the non-linear mechanical response of the vertex model (VM) of biological tissue under compression and dilation. The VM is known to exhibit a transition between rigid and fluid-like, or floppy, states driven by geometric incompatibility. Target perimeter and area set a target shape which may not be geometrically achievable, thereby engendering frustration. Previously, an asymmetry in the linear elastic response was identified at the rigidity transition between compression and dilation. Here we show and characterize how the asymmetry extends away from the transition point for finite strains. Under finite compression, an initially solid VM can totally relax perimeter tension, and thereby have reduced bulk and shear modulus. Conversely, an initially floppy VM under dilation can rigidify and have a higher bulk and shear modulus. These observations imply that re-scaling of cell area shifts the transition between rigid and floppy states. Based on this insight, we calculate the re-scaling of cell area engendered by intrinsic curvature and write a prediction for the rigidity transition in the presence of curvature.
The shift of the rigidity transition in the presence of curvature for the VM provides a new metric for predicting tissue rigidity from image data for curved tissues in a manner analogous to the flat case.
\end{abstract}

\maketitle


\section{Introduction}


Understanding the emergence of form in organ development presents a major challenge to current continuum physics modeling of biological tissues. Unlike passive materials, certain tissues may tune their mechanical response to applied strains and forces by modifying cell shape and thereby be rigid or floppy. In particular, cell shape as characterized by the shape index $s\equiv \frac{P}{\sqrt{A}}$ has been found to serve as a signal for the onset of a solid-liquid transition at constant density in epithelial tissues \cite{park2015unjamming}. 

One widely studied model of epithelial tissues is the vertex model (VM) which models the epithelium as a collection of vertices and edges in the 2D plane, and reduces the tissue's structural data to a polygonal tiling with possible edge tension. Unlike conventional spring network models which penalize deviations away from each edge length, the VM instead only sets a target cell area due to 3D bulk tissue incompressibility, along with terms capturing cell-cell edge adhesion and cell contractility\cite{farhadifar2009dynamics}, which constrain the cell's perimeter. Thus the VM is less constrained than a spring network, e.g. crystalline solids, and naturally engenders zeros modes for any polygonal tiling.\cite{hernandez2021geometric,huang2021shear,farhadifar2009dynamics}.

What's more these zero modes exist at the level of a single cell. For example, assuming all cell edges have identical adhesion and contractility, the VM energy reduces to penalizing harmonic deviations away from a target area $A_0$ and target perimeter $P_0$. 
%
%
Thus in the VM each cell has 2 shape constraints, but a general polygon has at least 3 degrees of freedom such as is the case for triangles \cite{hernandez2021geometric}. 

Based on constraint counting it seems the VM can never support a rigid state. Nonetheless the VM exhibits a rigidity transition between solid and floppy states tuned by the target shape index $s_0=\frac{P_0}{\sqrt{A_0}}$ about a critical point $s_0^*$ \cite{bi2015density}. The transition is due to a geometric constraint set by the isoperimetric inequality which gives a lower bound on the ratio of $\frac{P}{\sqrt{A}}$ for n-gons admissible on the plane \cite{osserman1978isoperimetric}.
\begin{align}
    \frac{P}{\sqrt{A}}\geq s_0^*(n)
\label{eq:isoperimetric}     
\end{align}
Where $s_0^*=\sqrt{4n\tan\left(\frac{\pi}{n}\right)}$ is the isoperimetric quotient.
The lower bound sets an incompatible regime $\frac{P_0}{\sqrt{A_0}}<s_0^{{*}}$ where polygons cannot simultaneously achieve $A_0$ and $P_0$, and a compatible regime $\frac{P_0}{\sqrt{A_0}}\geq s_0^{{*}}$ where polygons may achieve both $A_0$ and $P_0$. This geometric constraint on shape indicates that rigidity stems from self-tension due to geometrically incompatibility. 

For VM simulations consisting of ordered tilings (triangles, squares, hexagons) in the plane the rigidity transition occurs exactly at $s_0^*(n)$ \cite{hernandez2021geometric}. Whereas for disordered VM simulations the transition occurs at approximately $s_0^{*}(5)$ \cite{bi2015density}. 

The linear response of the VM to mechanical deformations is well studied \cite{farhadifar2007influence,hernandez2021geometric, staple2010mechanics,murisic2015discrete} but the non-linear response relatively less so. Recent work showed that the VM exhibits shear-thickening in the compatible regime \cite{huang2021shear}. In the same vein, this paper presents a careful study of the non-linear elasticity of the VM under finite dilation and compression via a mean-field approach and simulation. In previous work by the authors, the onset of compatibility in the VM at $s_0^*$ showed anomalous elasticity  as reflected by an asymmetric bulk modulus under dilation and compression, as well as coupling between stretching and shear modes \cite{hernandez2021geometric}. 

In this article, we show that the asymmetry of the bulk modulus extends away from $s_0^*$ under finite compression and dilation. In particular, the VM exhibits a dilation-hardening for compatible tissues and a compression-softening in incompatible tissues for finite critical strain. The hardening (softening) nonlinear response to dilation (compression) is reflected by a jump (drop) discontinuity of the bulk modulus and is associated with the sudden lifting (onset) of zero-modes.

The mechanism of a tissue fine-tuning their rigidity in response to areal re-scaling relates to several biological processes such as tissue growth, shrinkage, applied deformations, and in particular spontaneous generation of Gaussian (intrinsic) curvature. Based on insight from the planar 2D non-linear elasticity, we use of mean field theory to predict how local compression/dilation due to intrinsic curvature shifts the transition point between rigid and floppy states.

%
%


The organization of the paper is as follows: in section II  we define our mean-field approach of the VM which models 2D tissue elasticity at the single cell level. Section III outlines the calculation of the non-linear bulk modulus of the mean-field model and discusses simulation results. In section IV we present a Landau energy argument to elucidate the connection between the asymmetry of the bulk modulus and how a finite critical strain controls the onset/lifting of zero-modes. Section V uses our results from the mean field theory to predict the effective critical shape index for cells on a curved surface. We compare our prediction for the rigidity transition in the presence of curvature to simulations done by \cite{sussman2020interplay}. We conclude with discussion in section VI.
\section{Mean-Field Theory of ordered vertex model}
Our mean field theory for the VM is defined as a uniform regular 2D tiling with all cells responding identically to applied deformations. The tissue energy is the sum of all individual cells, and therefore our mean field approach reduces the VM to a single polygonal cell. All bulk tissue properties, such as elastic moduli, are defined at the single cell level response. Details of the mean field model are given in appendix C, and a thorough study by the authors is in \cite{hernandez2021geometric}. For concreteness, our simulations and mean field theory are for hexagonal cells unless stated otherwise. All results hold analogously for other polygons.

The tissue energy per cell is rescaled by $\kappa_A A_0^2$ such that the energy only has two dimensionless free parameters. The VM energy per cell is cast as
\begin{align}
E =  \frac{1}{2}\left(a-1\right)^2 +\frac{r}{2}\left(p-s_0\right)^2
\label{eq:EnergyVMrescaled} 
\end{align}
where $r\equiv \frac{\kappa_P}{\kappa_AA_0}$ is the rigidity ratio, and $s_0\equiv\frac{P_0}{\sqrt{A_0}}$ the target shape index, and $a$ and $p$ are the scaled area and perimeter. 

To parameterize cell shape degrees of freedom, we work with a Cartesian coordinate system $(X,Y)$ encompassing the cell with $Y$ along the height, and $X$ along the width. The area and perimeter of a cell are purely geometric objects, and their shape change under various deformations can be computed given a transformation law. 

\begin{figure}
    \centering
    \includegraphics[width=0.5\textwidth]{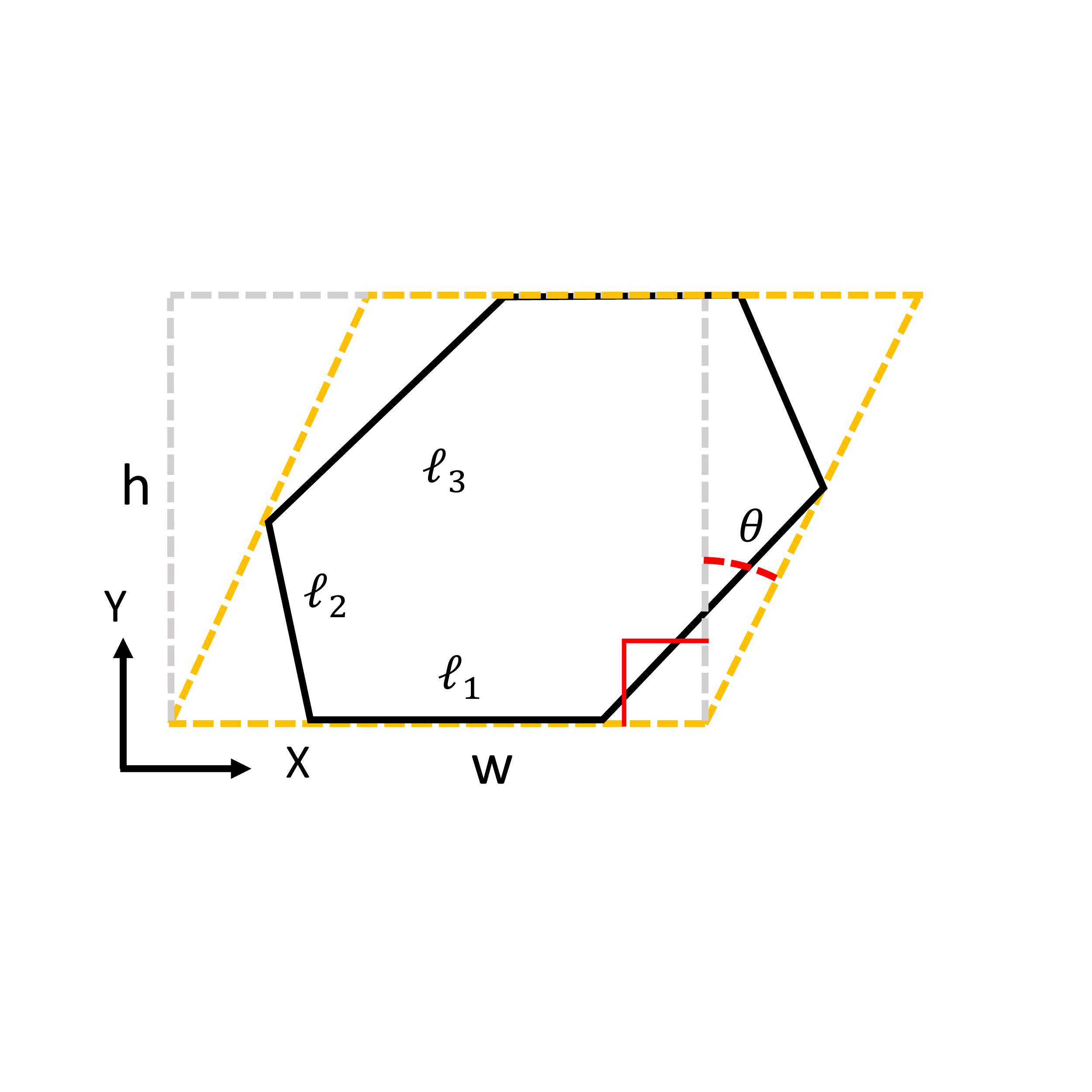}
    \caption{Under compression/dilation, the cell may respond via an affine self-shear transformation where it tilts either right or left. Only parallel edges change length in the same manner, whereas neighboring edges change length differently.}
    \label{fig:shearedhexagon}
\end{figure}

The externally imposed dilation and compression are implemented via an overall re-scaling of the cell's height $h$ and width $w$
via the transformation $w\rightarrow w(1+\epsilon)$, $h\rightarrow h(1+\epsilon)$, where  $\epsilon \in (-1,1)$. In response to the strain, the cell may also spontaneously shear while maintaining the imposed rescaled area,
as shown in Fig~\ref{fig:shearedhexagon}.
This "tilt" response is parametrized via a self-shear deformation, with $w\rightarrow w +t h$, $h\rightarrow h$, where $t(\theta)\equiv \tan\left({\theta}\right)$, and represents the shape degeneracy of cells.
%
In addition, hexagonal cells can also respond via non-affine transformations, and what's more these non-affine responses yield softer shear and Young's moduli in the incompatible regime~\cite{StaddonHernandez}. In this study we preclude non-affine pathways for cell response as our previous work showed this approximation well captures the response under imposed isotropic compression/dilation \cite{hernandez2021geometric}.

The deformed energy of an isotropically dilated or compressed cell is then given by 
\begin{align}
    E(\epsilon,\theta; s_0, r) =& \frac{1}{2} \left[\ell^2(1+ \epsilon)^2-1\right]^2 
    + \frac{r}{2}\left[p(\epsilon,t(\theta),\ell)-s_0\right]^2\;,
\end{align}
where $\ell(s_0,r)$ is the rescaled ground state characteristic cell size (see appendix C for details) and the deformed perimeter is 
\begin{align}
p(\epsilon,\theta,\ell)&= 
\frac{\sqrt{2}\ell}{3^{3/4}}(1+\epsilon)
\bigg(2\left(1+t\right) 
+\sqrt{1+(t(\theta)-\sqrt{3})^{2}} \nonumber\\ 
& +\sqrt{1+(t(\theta) +\sqrt{3})^{2}}
\bigg) \;.
\end{align}
%
%

If we set $\epsilon=0$ and minimize with respect to $\theta$ we recover the previous ground states results in \cite{hernandez2021geometric} where the ground state energy is gapped for $s_0<s_0^*$ and vanishes for $s_0\geq s_0^*$, populated by a manifold of degenerate shapes parametrized by $\theta$. To study the response to $\epsilon\neq 0$ we minimize $\theta$ as a function of applied strain in a manner analogous to our study of the linear response \cite{hernandez2021geometric}.  Formally, the energetic response is given by
\begin{align}
E = \min_{\theta} E(\theta;\epsilon,s_0,r)
\end{align}
Because height and width are fixed by dilation/compression the energy minimization is 1D and corresponds to solving,
\begin{align}
    \frac{\partial E}{\partial \theta}\bigg|_{s,r,\epsilon} = \left(p-s_0\right)\frac{\partial p}{\partial \theta}\bigg|_{s_0,r,\epsilon}=0
    \label{eq:analyticalminimization}
\end{align}
Which has two solutions: either a cell utilizes shape degeneracy via $\theta$ so that the perimeter accommodates both dilation/compression \textit{and} target shape index $s_0$, or the perimeter is totally set by dilation/compression with no tilt response. The relevant energy minimizing solution is a function of $s_0, r$ and $\epsilon$. 
\section{Nonlinear elasticity in the Euclidean Plane}

The non-linear response under finite dilation and compression is characterized by the bulk modulus, defined as 
\begin{align}
 K= \frac{1}{2 a_{\text{cell}}} \left(\frac{\partial^2 }{\partial \epsilon^2} \min_{\theta_{\text{min}}} E(\epsilon,\theta; s_0, r) \right)_{s_0,r,\epsilon}
 \label{eq:Bulkmodulusmeanfield} 
\end{align}
where $a_{\text{cell}} = \frac{3\sqrt{3}}{2} \ell^2$ is the rescaled cell area. Evaluating Eq. \ref{eq:Bulkmodulusmeanfield} at $\epsilon=0$ yields the linear response, whereas a finite $\epsilon$ gives the non-linear response under finite strains. 

It is crucial that the minimization occurs before differentiation because the self-shear response $\theta$ is implicitly dependent on $\epsilon$ via Eq. \ref{eq:analyticalminimization}.

In the incompatible rigid state, ($s_0<s_0^*$), we find that the mean field model and simulation exhibit a discontinuous drop in the bulk modulus at a critical compression. The discontinuity occurs due to a spontaneous self-shear of the cell which allows perimeter tension to vanish. Conversely, under dilation the bulks modulus remains continuous as a function of strain, see Fig. 2 Plot A. Larger choices of $r$ will shift the critical strain up, reflecting how a higher perimeter tension may support higher compression before giving way to spontaneous self-shear.  

In the compatible floppy state,  ($s_0>s_0^*$), the bulk modulus is continuous under any finite compression but discontinuous jump at some critical dilation, see Fig. 2 plot B. At sufficient dilation, the  zero-modes of the degenerate ground state are "exhausted", resulting in a frustrated and thereby rigid state. Unlike the incompatible state, the critical dilation is insensitive to $r$. 

At the transition, $s_0^*$, both dilation hardening and compression softening are present for arbitrarily small strains, and reflect an asymmetry of the response to area re-scaling. For various dilation/compression strain magnitudes we plot the difference between dilation and compression bulk modulus $ \Delta K$ vs. $s_0$ as a measure of asymmetry of the response in Fig. 2 plot C. Around the critical shape index, $s_0^*$, the asymmetry extends continuously away from the critical point for even modest strain values ($>0.002$). Note the curve $\Delta K$ vs. $s_0$ is also not left-right symmetric along the $s_0=s_0*$ axis; this is due to the critical strain only depending on the choice of $r$ in the incompatible state but not in the compatible state. 

The origin of the bulk modulus discontinuity can be partial elucidated by writing out Eq.8 explicitly
\begin{align}
    K =\frac{1}{2 a_{\text{cell}}} \left(a^2 + r\big(p(\theta_{min}) - s_0\big)\frac{\partial^2 p(\theta_{min})}{\partial \epsilon^2} + r\left(\frac{\partial p(\theta_{min})}{\partial \epsilon}\right)^2\right).
\end{align}
In the incompatible state, if cells can accommodate target perimeter and imposed strains simultaneously, the perimeter tension, $r\big(p(\theta_{min}) - s_0\big)$, vanishes thereby resulting in a discontinuous drop of the bulk modulus. Conversely, in the compatible state sufficient dilation will yield sudden contribution from perimeter tension.
In the following section, we apply a Landau-type energy analysis to understand the trigger of shape zero-modes under dilation and compression. 
\section{Landau energy expansion}
To understand how compression, or dilation, may trigger, or lift, shape degeneracy we treat $\theta$ as an order parameter for the onset of shape degeneracy such that $\theta \neq 0$ when cells adjust their shape to accommodate imposed strains, and $\theta= 0$ when cells remain rigid (no cell response). 

Before minimization, we expand the energy in power of $\theta$ to quartic order,
\begin{align}
    E(\epsilon,\theta;s_0,r)\approx E_0+ \frac{\alpha}{2} (\epsilon,s_0,r) \theta^2 +\frac{\beta (\epsilon,s_0,r)}{4} \theta^4 + \mathcal{O}(\theta^6)
\end{align}
Where
\begin{align}
    \alpha &=\frac{3^\frac{5}{4}}{2\sqrt{2}}r\ell(1+\epsilon) \bigg(s_0^*(6)\ell(1+\epsilon)-s_0\bigg),
    \label{eq:landauparameter} \\
    \beta &= \frac{3}{128}3^{\frac{1}{4}} r \ell (1+\epsilon)\bigg(13\sqrt{2}s_0-28\times 3^{\frac{1}{4}}(1+\epsilon)\ell\bigg)
\end{align}
For the quartic approximation energy minimization yields $\theta_{\text{min}}=0$ for $\alpha>0$, or $\theta_{\text{min}}=\pm \sqrt{\frac{-\alpha}{\beta}}$ for $\alpha<0$. 
The Landau expansion highlights the role of strain $\epsilon$ as tuning parameter between the cell responding with $\theta_{\text{min}}=0$ or by spontaneously tilting via a shear of $\theta_{\text{min}} =\pm \sqrt{\frac{-\alpha}{\beta}}$. 
The form of $\alpha$ reflects an asymmetric response between compression versus dilation. From $\alpha$ we can extract the critical strain, $\epsilon_*$, which controls the onset/lifting of shape degeneracy. 
\begin{align}
\epsilon_{*}=\frac{1}{\ell}\frac{s_0}{s_0^{*}(6)}-1
\label{eq:criticalstrain} 
\end{align}
The critical strain vanishes at the critical shape index and coincides with the the failure of linear elasticity for any applied strain \cite{hernandez2021geometric}. In the compatible regime $\ell=1$ because target area is always achieved and $\epsilon_*$ is independent of rigidity ratio $r$. Whereas in the incompatible $\epsilon_{*}$ is dependent on $r$ through $\ell$. 

We input the cell response via $\theta_{\text{min}}$ into the energy and expand in powers of strain $\epsilon$. 
\begin{align}
E(\epsilon;s_0,r)=& \min_{\theta} E(\epsilon, \theta; s_0,r)\\
\approx& \min_{\theta} E_0+ \frac{\alpha}{2} (\epsilon,s_0,r) \theta^2 +\frac{\beta (\epsilon,s_0,r)}{4} \theta^4 + \mathcal{O}(\theta^6)\\
=& E_0 +  \frac12 \left(\frac{\partial^2 E (\theta_{\text{min}})}{\partial \epsilon^2}\right)\epsilon^2 + \mathcal{O}(\epsilon^3)    
\end{align}
In the final line the harmonic coefficient contains contributions from $\theta$ which reduce the overall response of the tissue. If we did not minimize over $\theta$ before expanding in $\epsilon$, the resultant deformed energy would not reflect the self-shear response due to cell shape changes.

%
%

\label{sec:} 
 \begin{figure*}
	\centering
	\includegraphics[width=0.9\linewidth]{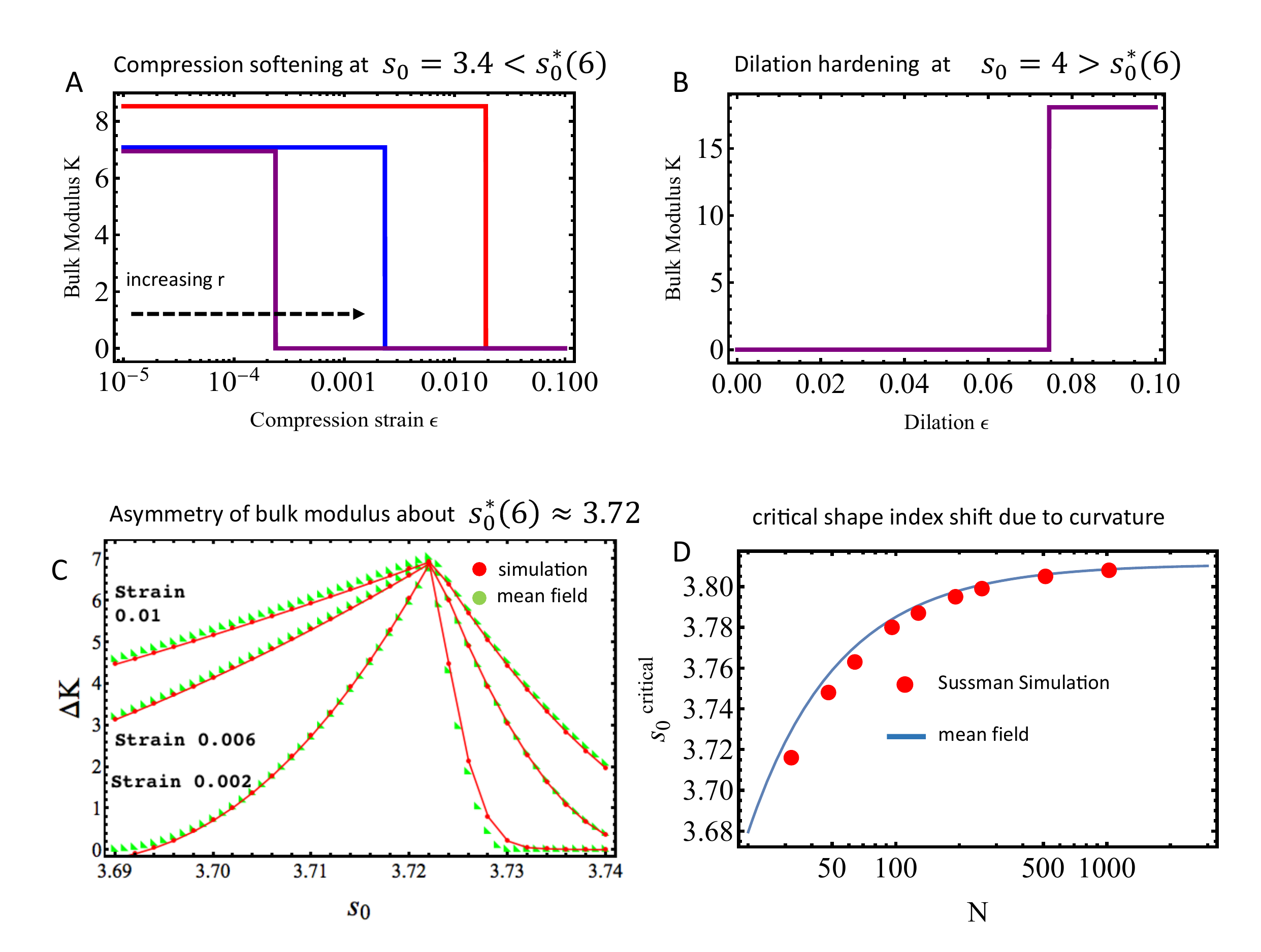}
	\caption{Plot A shows the mean field non-linear response bulk modulus versus compression in the incompatible regime. At a critical strain $\epsilon_*$, there is sudden discontinuous softening. The critical strain depends on rigidity ratio, and we plot curves for $r=1,10,100$. In the compatible regime, plot B shows the mean field bulk modulus hardening at a critical compression. Note that $\epsilon_*$ in the incompatible regime is dependent on $r$ and thereby sensitive to the balance between perimeter and areal elasticity. Whereas in the compatible regime, the critical strain is only a function of shape index. Plot C plot the difference in the linear response between dilation and compression, and shows the continuous extension of the response asymmetry extends away from the critical shape index. Plots A,B, and C corresponds to hexagons. Plot D shows the effective critical shape index for a random tiling of N cells on a sphere of radius $L=\sqrt{\frac{N}{4\pi}}$. The mean field prediction is for pentagons , whereas the simulation is for a disordered VM taken from \cite{sussman2020interplay}.} 
	\label{fig:}
\end{figure*}
%
%
A summary of the consequences of $\theta$ on the response are as follows:
In the incompatible regime, the $\theta_{\text{min}}=0$ solution corresponds to the linear response in the solid state (see Fig. 3), whereas $\theta_{\text{min}}=\pm \sqrt{\frac{-\alpha}{\beta}}$ corresponds to the softer renormalized nonlinear response at critical compression strain. On the other hand, in the compatible regime, the linear response is always given by $\theta_{\text{min}}=\pm \sqrt{\frac{-\alpha}{\beta}}$ which allows perimeter tension to vanish. The hardening under finite dilation occurs at a critical dilative strain $\epsilon_*$ and corresponds to a switch from $\theta_{\text{min}}=\pm \sqrt{\frac{-\alpha}{\beta}}$ to $\theta_{\text{min}}=0$, resulting in a higher response. This hardening phenomena is due to the cell's inability access degenerate ground states to accommodate large dilation. 

\subsection*{Calculating new critical shape index }
So far our mean-field model has predicted how compression/dilation controls the onset/lifting of shape degeneracy. If one defines the rigidity of a solid by the availability of cell level zero modes, than the mean field treatment suggests that dilation and compression shift the rigidity transition of the VM.

The precise shift of the critical shape index can be found by solving $\alpha=0$ for $s_0$ and setting the strain, $(1+ \epsilon)$, as a material parameter for local cell dilation/compression. Then solving for $s_0$ defines a "new" critical shape index as set by dilation/compression. The effective critical point $s_0^{\text{critical}}$ is
\begin{align}
\alpha &= \bigg(s_0^*(n)\ell(1+\epsilon)-s_0^{\text{critical}}\bigg)=0 \nonumber\\
&\implies s_0^{\text{critical}} = s_0^*(n)\ell(1+\epsilon)
 \label{eq:curvedcriticalshape} 
\end{align}
Where we used the modified version of Eq. \ref{eq:landauparameter} for n-gons (see appendix C).
Of course this is not a complete shift of the rigidity transition since even if perimeter tension vanishes there will always be areal tension. Hence the system remains energetically frustrated, albeit less so, but soft. Furthermore, $\epsilon$ needs to be determined by some cell growth/compression process. 

If cells are modeled on a 2D surface VM consisting of a 2D tiling embedded in 3D then cells may accommodate in-plane dilation/compression by buckling and achieve true geometric compatibility in a curved geometry. For example, geometrically frustrated (rigid) cells under compression may buckle to achieve target area and perimeter by utilizing additional d.o.f. associated with curvature, and thereby become compatible. By the same token, geometrically compatible (floppy) cells experiencing dilation/growth will eventually rigidity and become frustrated but may buckle to find a more optimally/energetically lower frustrated state. Both case correspond to cells utilizing curvature-related d.o.f to satisfy area and perimeter constraints and thereby fine tuning their rigidity. 

In support of this perceptive, simulations have reported that the rigidity transition - as signalled by the critical target shape index - is indeed sensitive to curvature. For example, a VM constrained on a sphere \cite{sussman2020interplay} has a downward shifted $s_0^*$ dependent on curvature magnitude and relative cell size.

In the rest of this article, we calculate the effective dilation/compression set by curvature by calculating cell area on a curved surface relative to its flat counterpart. Utilizing Eq.\ref{eq:curvedcriticalshape}, we predict the target shape index on curved 2D tissues.  
\section{Rigidity transition in the presence of curvature.}
%
%
We extend the mean field calculation by pertubatively calculating expressions for cell area on surfaces of constant curvature in powers of $K_{G}R_{\text{cell}}^2$, where $K_{G}$ is the Gaussian curvature, and $R_{\text{cell}}$ the cell radius. Polygons and discs of fixed radius have different total area depending on the surface on which they are inscribed. The effective strain stems from the mismatch in area of curved cells from their flat analog. 

All geometric information of a surface $M$ is encoded in the metric tensor $\mathbf{g}$ and we denote a surface generically as $(\mathbf{g},M)$. 
For a general shape/cell on a surface, $D\subset M$, the area is defined by
\begin{align}
A &= \int_{D}\sqrt{\det g} d^2x
\end{align}
Note that the determinant of the metric serves as a weight which accounts for the local compression/dilation of the distances between points.
Unlike the planar setting of the mean field model where $g_{ij}=\delta_{ij}$, the metric on curved surfaces - even uniformly curved - is not homogeneous but a function of space. 
Nonetheless, the metric always admits a local pertubative expansion in \textit{normal coordinates}, $(x_0,x^i)$ which are defined by the condition that geodesics can be locally parameterized as straight lines, i.e. $\gamma(\lambda)=(x^1\lambda, x^2\lambda)$. In these coordinates, a pertubative expansion of the metric in powers of curvature yields
%
%
%
\begin{align}
   \det (g) = 1-\frac{1}{3} \mathcal{R}|x|^2  +\mathcal{O}(|x|^3) 
\end{align}
Where $\mathcal{R}$ is the scalar curvature which is twice the Gaussian curvature $K_G=2\mathcal{R}$. 
The expansion reflects how variations of the metric are tied to curvature, and is locally approximated as flat with corrections due to curvature.
An outline of the derivation for Eq. 18 is given in appendix A.

%
%
%
%

In our calculation we restrict attention to surfaces of uniform curvature by fixing $\mathcal{R}$, and hence we only consider uniformly flat, sphereical, and saddle-like surfaces. Of course, real curved biological tissues are not not uniform and might have a boundary. Our approximation is controlled by the dimensionless number set by the radius $R_{\text{cell}}$ of the cell over the radius of curvature $L_{K}$
\begin{align}
C\equiv\frac{R_{\text{cell}}}{|L_{K}|}
\end{align}
where $K_{G} \equiv \pm \frac{1}{L_{K}^2} $. 
Our mean field result will hold best for tissues with moderate curvature or relativity small cells. 
\section*{Computing area}
Using the metric expansion the area is expanded to quadratic order, 
\begin{align}
A &=\int_{D} \sqrt{\det g} d^2x \nonumber \\
&= \int_{D} d^2 x -  \frac{\mathcal{R}}{3}\int_{D} |x|^2 d^2x+ \text{h.o.t.}
\end{align}
The first term yields the flat case. We generalize our calculation for n-sided polygons for easy comparison of various tilings. 

To parameterize the polygonal n-sided cell $D$ we decompose it into 2n triangles about the centroid as illustrated in figure 3. 
Details of the calculation are given in appendix B.
To quadratic order the area is 
\begin{align}
A = \Bar{A}\left(1- \frac{\mathcal{R}}{6}R_{\text{cell}}^2 f(n) +\mathcal{O}(R_{\text{cell}}^4) \right) 
\label{eq:curvedarea} 
\end{align}    
Where $f(n)\equiv \cos^2\left(\frac{\pi}{n}\right)\left(\frac{2}{3}+\frac{1}{3}\sec^2\left(\frac{\pi}{n}\right)\right)$, and $\Bar{A}=nR_{\text{cell}}^2 \cos\left(\frac{\pi}{n}\right)^2 \tan\left(\frac{\pi}{n}\right)$ is the flat area. In the limit of either very small cell size or very small curvature, we get the typical planar area. Morever, in the limit $n\rightarrow \infty$ the first correction yields $\frac{\mathcal{R}}{6}R_{\text{cell}}^2 \pi$ which reproduces the classical result of Bertand-Diguet-Puiseux on the area comparison of a 2D geodesic ball of radius $R_{\text{cell}}$ to its flat counterpart \cite{gray1979riemannian}. 
%
%
\section*{Rigidity transition shift}
From Eq.\ref{eq:curvedarea} (and using Eq. 3) we write down the induced dilation/compression strain set by curvature 
\begin{align}
\epsilon_{\mathcal{R}}(n)&=-1+ \sqrt{1-\frac{\mathcal{R}}{6}R_{\text{cell}}^2 f(n)}\nonumber\\
&=-1+\sqrt{ 1-\cos^2\left(\frac{\pi}{n}\right)\left(\frac{2}{3}+\frac{1}{3}\sec^2\left(\frac{\pi}{n}\right)\right)\frac{\mathcal{R}}{12} R^2_{\text{cell}}}
\end{align}
The effective strain depends on the number of edges due to the discrete rotational symmetry of polygons: the furthest edges are weighted differently than the midpoint of edges. This vanishes in the $n=\infty$ limit as the cell is approximates a disc and regains full rotational symmetry. 

%
%
The predicted shift in the critical shape index is
\begin{align}
s_0^{\text{critical}}(n,\mathcal{R}) &= s_0^*(n)\ell(n)(1+\epsilon_{\mathcal{R}}(n)) \\ \nonumber
&\approx s_0^*(n)(1+\epsilon_{\mathcal{R}}(n))
\end{align}
Where we've set $\ell \approx 1$, which restricts our prediction near the planar critical point or large rigidity ratio $r>>1$.


%
\section*{Comparison with simulation}
In the work by Sussman \cite{sussman2020interplay} a disordered vertex model of $N$ cells on a uniform sphere of radius $L=\sqrt{\frac{N}{4\pi}}$ was simulated for various $N$. At the onset of rigidity, the observable shape index per cell $\frac{P}{\sqrt{A}}$ was extracted and used to compute a numerical probability distribution. The peak of this distribution was used to define the critical shape index on the sphere. For large systems, the critical point was found to be sharply peaked at around $s_0^{*}\sim 3.812$, which corresponds to the the critical shape index for pentagons\cite{bi2016motility}. 


To compare with \cite{sussman2020interplay} we re-cast $\mathcal{R}R_{\text{cell}}^2$ in terms of $N$. The radius of curvature for a sphere gives the Gaussian curvature $K_{G}\equiv \frac{1}{L^2}$, which is 1/2 the scalar curvature. 
\begin{align}
 s_0^{\text{critical}}(n=5,K_G) \approx 3.812 \left(1-\frac{(7+\sqrt{5})}{72}\frac{2R_{\text{cell}}^2}{L^2}\right)^{1/2}
\end{align}
Sussman works on a sphere of raduis $L=\sqrt{\frac{N}{4\pi}}$ and average cell area is set to unity, i.e. $A_{\text{cell}}\equiv\frac{N}{4\pi L^2}=1$. We neglect cell packing since our mean field calculation is for a single cell, and so we take $R_{\text{cell}}$ as given for a pentagon. Thus $5R_{cell}^2 \cos\left(\frac{\pi}{5}\right)^2 \tan\left(\frac{\pi}{5}\right)=1 \implies $ $R_{\text{cell}}^2\approx \frac{1}{2.377} $. 

Therefore the relative ratio of cell size to radius of curvature goes as $\frac{R_{\text{cell}}^2}{L^2} \approx \frac{4\pi}{2.377}\frac{1}{N}$, yielding
\begin{align}
s_0^{\text{critical}}(n=5,K_G) \approx 3.812 \left(1-\frac{(7+\sqrt{5})}{36}\frac{4\pi}{2.377 }\frac{1}{N}\right)^{1/2}
\label{eq:pentagonshift}
\end{align}
%
%
%
%
Comparison to simulation data is shown in Fig. 2 Plot D. 

Besides expanding to higher order the calculation can be improved by computing the ground state characteristic cell size $\ell_0$ for curved vertex models, but this calculation is beyond our mean field approach. Additionally, \cite{sussman2020interplay} reports that the shape index distribution broadens for larger $R^2_{\text{cell}}K_G$, reflecting a greater diversity of polygons at the rigidity transition than the flat counterpart. Taking into account this greater diversity could help refine the curvature correction in Eq. \ref{eq:pentagonshift}. In particular, for relative large curvature other polygonal shapes besides the pentagon could more relevant for disordered systems. 
\section{Discussion}
Utilizing a mean field model we show the asymmetry of the linear response in the vertex model under dilation and compression extends away from the critical shape index for finite strains. The asymmetry reflects how an initially rigid tissue may be sufficiently compressed to induce shape degeneracy and thereby relax perimeter tension, yielding a softer bulk modulus. Furthermore, sufficient dilation applied to a compatible (floppy) cell lifts shape degeneracy, yielding an uptick in the bulk modulus. Thus applied dilation and compression shift the rigidity of the VM in 2D. 

Using this insight, we extend our mean field theory to calculate the effective dilation/compression engendered by intrinsic curvature and predict the precise shift in the rigidity transition by calculating the effective critical shape index. We compare our result to simulation done by \cite{sussman2020interplay} and find good agreement. 

Our mean field prediction provides a metric which can be applied to studying the rigidity transition in curved biological tissues in a manner analogous as the planar shape index.  

\section{Appendices}
\subsection{Details about metric expansion}
The purpose of this appendix is to give a brief explanation of the series expansion of the metric in terms of curvature. A more complete and rigorous treatment may be found in many textbooks on Riemannain geometry such as \cite{gray2003tubes, Willmore}, and in particular we follow the classic treatise \cite{veblen}. 

In general the metric is a second order tensor whose components are spatially dependent function of the surface. It governs all geometric data in that the distance between any two points is given by the line element
\begin{align}
ds^2 = g_{ij}(x)dx^idx^j   . 
\end{align}
About a given point $x_0\in M$ in some neighborhood, the components of the metric are approximated as constants to 1st order. One may diagonalize this linear approximation such that the metric at $x_0$ is given by $\delta_{ij}$. However to 2nd order the metric's components need not be also constants. In fact, if there exist coordinates such that the metric's Taylor series is constant up to 2nd order then the metric is totally flat in the neighborhood.

Normal coordinates about a point $x_0$ are defined locally where geodesics may be parametrized in local coordinates $x^i$ by $\lambda$ such that $\gamma^i(\lambda)= x^i \lambda$, where $\gamma(0)\equiv x_0$. 
The Christoffel symbols are extracted from the geodesic equation
\begin{align}
0=\frac{d^2 \gamma^{i}}{d\lambda^2} + \Gamma_{k\ell}^i \frac{d\gamma^k}{d\lambda} \frac{d\gamma^{\ell}}{d\lambda}    
\end{align}
%
Utilizing normal coordinates, Eq. 29 immediately implies $\Gamma_{k\ell}^i(x_0)=0$. Differentiation and index manipulation also yields the differential constraint equation.
\begin{align}
\partial_{j}\Gamma_{k\ell}^i(x_0)+\partial_{k}\Gamma_{\ell j}^i(x_0)+\partial_{\ell}\Gamma_{j k}^i(x_0)=0
\label{eq:diffconstraint}
\end{align}
The Riemann curvature tensor is defined as
\begin{align}
R^i_{jkl}&=\partial_k\Gamma^i_{jl}-\partial_l\Gamma^i_{jk} + \Gamma^i_{pk}\Gamma^p_{kl}+ \Gamma^i_{pl}\Gamma^p_{kj}
\end{align}
From the differential constraint and the definition of $R^i_{jkl}$, one can show
\begin{align}
&\partial_{l}\Gamma_{ij}^k  =-\frac{1}{3}\left(R_{ijl}^k+R_{jil}^k\right)
\end{align}
Symmetry of the metric implies the covariant derivative of the metric vanishes,i.e. $\nabla \mathbf{g}=0$ $\implies \partial_{k}g_{ij} -\Gamma^p_{jk}g_{ip} -\Gamma^p_{ik}g_{jp}=0$.
The second derivative of the metric in thes coordinates is
\begin{align}
\partial_{kl}^2g_{ij} = -\frac{1}{3}  \left(R_{klij}+R_{jlik}\right), 
\end{align}
The Taylor expansion of the metric in normal coordinates yields
\begin{align}
g_{ij}&= \delta_{ij}-\frac{1}{3}R_{ijk\ell}x^k x^{\ell} +\mathcal{O}(|x|^2)
\end{align}
Higher order terms can be generated iteratively by calculating higher order differential constraint equations from Eqs.\ref{eq:diffconstraint} and $\nabla \mathbf{g}=0$.

For 2D surfaces the Riemann curvature tensor only has a single d.o.f. and admits the representation \cite{Willmore}
\begin{align}
R_{iklj}= \mathcal{R} (g_{ik}g_{lj}-g_{ij}g_{kl})    
\end{align}
Where $\mathcal{R}=2K_G$ is the Ricci scalar curvature.
From this the Ricci tensor follows $R_{ij}\equiv g^{kl} R_{iklj} = \mathcal{R}g_{ij}$. Using the expansion of the metric, we have to lowest order
\begin{align}
R_{iklj}&= \mathcal{R} (\delta_{ik}\delta_{lj}-\delta_{ij}\delta_{kl}) + \mathcal{O}(|x|^2)   \\
R_{ij}& = \mathcal{R}\delta_{ij}  + \mathcal{O}(|x|^2)
\end{align}
These expressions reflect that locally any surface looks either flat $(\mathcal{R}=0)$, spherical $(\mathcal{R}>0)$, or saddle-like $(\mathcal{R}<0)$.

To lowest order the metric expansion about $p$ becomes 
\begin{align}
g_{ij} =&\delta_{ij} - \frac{1}{3} \mathcal{R}   (\delta_{ik}\delta_{lj}-\delta_{ij}\delta_{kl}) x^k x^l +\mathcal{O}(|x|^3) \\
\end{align}
and determinant yields 
\begin{align}
   \det (g) = 1-\frac{1}{3} \mathcal{R}|x|^2  +\mathcal{O}(|x|^3) 
\end{align}
which shows explicitly how curvature induces local compression  or dilation. Higher order terms contain gradients and higher order invariants of $R_{ijkl}$, and therefore are completely determined by $K$. Thus it follows that if the quadratic contribution vanishes, then the metric is totally flat in the neighborhood. 
\subsection{Pertubative polygon area expansion}
To explicitly parameterize the polygonal cell $D$, we will consider a regular n-gon and decompose it into 2n-triangles about its centroid as pictured in figure 3. 

Working in terms of polar coordinates, this yields for the first term
\begin{align}
\int_{ D} d^2x &= 2n \int_{0}^{\frac{\pi}{n}}\int_0^{R_{\text{cell}}\cos(\frac{\pi}{n})\sec\theta} d\theta rdr \nonumber\\
&= nR_{\text{cell}}^2 \cos\left(\frac{\pi}{n}\right)^2 \tan\left(\frac{\pi}{n}\right)
\end{align}
In the limit of $n\rightarrow \infty$ we get $\pi R_{\text{cell}}^2$, as expected for circles. 
Using the same coordinate system, we compute the first correction due to curvature
\begin{align}
\frac{\mathcal{R}}{3}\int_{ D} |x|^2 d^2x &= \frac{\mathcal{R}}{3} 2n 
\int_{0}^{\frac{\pi}{n}}\int_0^{R_{\text{cell}}\cos(\frac{\pi}{n})\sec\theta} d\theta r^3dr
\nonumber\\    
&= \frac{\mathcal{R}}{6}R_{\text{cell}}^4 \cos^4\left(\frac{\pi}{n}\right) n \left(\frac{2}{3} +\frac{1}{3}\sec^2\left(\frac{\pi}{n}\right)\right)\tan \left(\frac{\pi}{n}\right)
\end{align}
\subsection{Mean field vertex model}
The mean field model is defined by the area and perimeter of a single cell, which is parameterized by n-edges $\nu_{\alpha}$ given by
\begin{align}
  \vec{\nu}_{\alpha}\equiv  \ell_0 \left(\cos\left(\frac{2\pi \alpha}{n}\right),\sin\left(\frac{2\pi \alpha}{n}\right)\right)
\end{align}
Where $\ell_0$ the characteristic cell edge length. 
The perimeter is the sum of each edge length
\begin{align}
    P =& \sum_{\alpha}^n \sqrt{\vec{\nu}_{\alpha}\cdot \vec{\nu}_{\alpha}} 
\end{align}
Under an affine transformation, denoted as the matrix $\mathbf{F}$, the perimeter transformations as
\begin{align}
    P =& \sum_{\alpha}^n \sqrt{(\mathbf{F}\vec{\nu}_{\alpha})\cdot (\mathbf{F}\vec{\nu}_{\alpha}}) 
\end{align}
The area can be calculated by the cross product
\begin{align}
A =\int_{D} dx^2 = n | \vec{a}\times \vec{b}|    
\end{align}
where $\vec{a}$ and $\vec{b}$ are defined in Fig.3. 

The polygon will be changed under some affine transformation $\mathbf{F}$. The area term is striaigtforeward since cross product transforms as $| (\mathbf{F}\vec{a})\times (\mathbf{F} \vec{b})|= \det(\mathbf{F}) | \vec{a}\times \vec{b}|$. Thus under any affine transformation, the area term takes the simple form 
\begin{align}
A =& \det(\mathbf{F}) n | \vec{a}\times \vec{b}|   \\
 =&\det(\mathbf{F}) \frac{n}{4}\ell_0^2 \cot\left(\frac{\pi}{n}\right)
\end{align} 
\begin{figure}
	\centering
	\includegraphics[width=.5\textwidth]{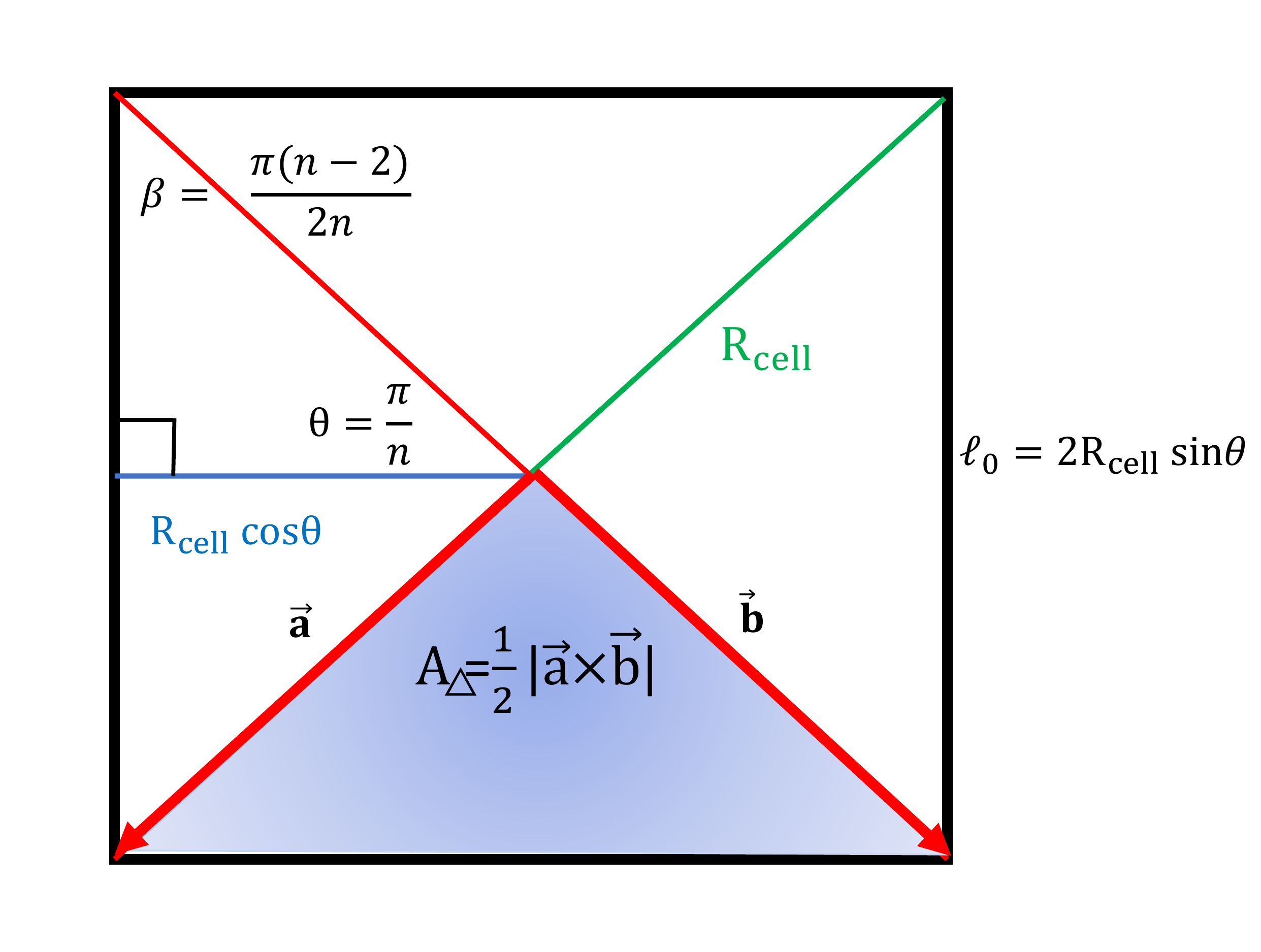}
	\caption{To make our calculation easily applicable to squares, pentagons, and hexagons we work with this figure to switch between the different polygons and different choices of encoding cell shape either via edge length $\ell_0$ or cell radius $R_{\text{cell}}$.} 
	\label{fig:}
\end{figure}
The energy per cell is cast as
\begin{align}
 E = \frac{\kappa_A}{2} \left(\frac{n}{4}\ell_0^2 \cot\left(\frac{\pi}{n}\right) \det (\mathbf{F})-A_0\right)^2  \nonumber\\ +\frac{\kappa_P}{2}\left(\sum_{\alpha}^n \sqrt{(\mathbf{F}\vec{\nu}_{\alpha})\cdot (\mathbf{F}\vec{\nu}_{\alpha}})-P_0\right)^2  
\end{align}
To non-dimensionalize we define reference lengths $\ell_A$ and $\ell_P$ such that
\begin{align}
A_0&= \frac{n}{4}\ell_A^2 \cot\left(\frac{\pi}{n}\right)   \\
P_0&= n\ell_P
\end{align}
And rescale energy by $\kappa_A A_0^2$, yielding
\begin{align}
E = \frac12 \left(\ell^2\det (\mathbf{F})-1\right)^2+\frac{r}{2}\left(\ell\sum_{\alpha}^n \sqrt{(\mathbf{F}\vec{\nu}_{\alpha})\cdot (\mathbf{F}\vec{\nu}_{\alpha}})-s_0\right)^2
\end{align}
Where $r\equiv\frac{\kappa_P}{\kappa_A A_0}$, $s_0\equiv \frac{P_0}{\sqrt{A_0}}$ is the target shape index, and $\ell\equiv\frac{\ell_0}{\sqrt{A_0}}$ is the re-scaled characteristic cell edge length. 

In the incompatible state, the ground state corresponds to a regular polygon with $\ell_0$ defined to minimize the energy. This involves solving the following cubic equation.
\begin{align}
\frac{\partial E}{\partial \ell}=0
\end{align}
The relevant solution obeys the inequality $\ell\leq 1$ for all $s_0\leq s_0^*$.  In the compatible state energy minimization yields that $\ell\equiv 1$ for choices of $r$ and $s_0\geq s_0^{*}$. 
\subsubsection*{Inputting deformations}
The linear transformation $\mathbf{F}$ encodes both applied deformations and cell response. We assume that all shape distortions of the cell can be captured in by linear affine transformation. Although the non-affine contribution is important in calculating exactly even the linear response of the vertex model as shown by the authors \cite{StaddonHernandez}. 

The applied deformation is set by a transformation rule. In this article we focus on compression and dilation, which yields the matrix
\begin{align}
 \mathbf{F}_{\epsilon} =
\begin{pmatrix}
1+\epsilon & 0\\
0 & 1 +\epsilon
\end{pmatrix} 
\end{align}
On top of this, we also will allow the cell to adjust its perimeter without changing the imposed re-scaled area. This imposes the constraint 
\begin{align}
\det (\mathbf{F}^{\text{cell}})=1    
\end{align}
Which only fixes a single degree of freedom, leaving in principle three components of $\mathbf{F}_{\text{cell}}$ free. For simplicity, we only consider the the cell's response by tilting through a simple shear transformation
\begin{align}
 \mathbf{F}^{\text{cell}}_{\theta} =
\begin{pmatrix}
1 & \tan(\theta)\\
0 & 1 
\end{pmatrix}    
\end{align}
Note that our mean field model is invariant under rotations since area and perimeter are rotationally invariant objects. Additionally, if we include other affine transformation, all their respectively matrices $\mathbf{F}$ will commute with one another.

We set the overall net deformation gradient in the mean field model as
\begin{align}
\mathbf{F} =\mathbf{F}^{\text{cell}}_{\theta}\cdot \mathbf{F}_{\epsilon}=
\begin{pmatrix}
1 & \tan(\theta)\\
0 & 1 
\end{pmatrix}    
\begin{pmatrix}
1+\epsilon & 0\\
0 & 1 +\epsilon
\end{pmatrix} 
\end{align}
With these two shape degrees of freedom, $\epsilon$ and $\theta$, we study the non-linear response to finite compression/dilation and extract our rigidity transition shift due to curvature. 
%



\bibliography{apssamp}

\end{document}